\documentclass[12]{article}
\usepackage{amsmath,mathrsfs}
\usepackage{authblk}
\usepackage{alphalph}
\usepackage{array}
\usepackage{graphicx}
\usepackage{epstopdf} 
\usepackage{color}
\graphicspath{{./Figures/}{./Figures/Simulations/}{./Figures/Noise/}{./Figures/Experiment/}{./Figures/overlap/}{./Figures/comparison/}}
\usepackage{float}
\usepackage{caption}
\usepackage{subcaption}
\usepackage[colorinlistoftodos,prependcaption,textsize=10]{todonotes}

\title{Lateral position correction in ptychography with sub-pixel accuracy}
\date{}
\author{P. Dwivedi, A.P. Konijnenberg, S.F. Pereira, and H.P. Urbach}
\affil{Optics Research Group, Department of Imaging Physics, Delft University of Technology, Lorentzweg 1, 2628CJ, Delft, Netherlands}

\begin{document}
\maketitle

\begin{abstract}
Ptychography, a form of Coherent Diffractive Imaging, is used with short wavelengths (e.g. X-rays, electron beams) to achieve high-resolution image reconstructions. One of the limiting factors for the reconstruction quality is the accurate knowledge of the illumination probe positions. Recently, many advances have been made to relax the requirement for the probe positions accuracy. Here, we analyse and demonstrate a straightforward approach that can be used to correct the probe positions with sub-pixel accuracy. Simulations and experimental results with visible light are presented in this work. 
\end{abstract}

\section{Introduction}
Coherent Diffractive Imaging (CDI) is a lensless imaging technique which uses far-field diffraction intensity patterns to reconstruct the image of an object. Ptychography is a form of CDI, where multiple far-field diffraction patterns corresponding to overlapping illuminated regions of the object are collected, and the object is reconstructed \cite{Ptycho2}. For the reconstruction of the object, the Ptychographical Iterative Engine   (PIE) \cite{PIE} is used of which many different variants have been developed \cite{ePIE, fPIE, probe_PIE, HIO-PIE}. PIE has been found to be robust if the a priori information such as the illumination probe function and the lateral probe positions are accurately known \cite{tolerance}. Several methods exist which can overcome the requirement for the accuracy of the a priori information. For example, Extended PIE (ePIE) can reconstruct the object as well as a poorly defined probe function \cite{ePIE}. However, ePIE has been found to be sensitive to the probe positioning errors, especially in applications involving short wavelengths such as X-rays and electron beams \cite{limitations}. For these short wavelengths, the required accuracy in the probe positions should be in some cases of the order of 50 pm \cite{ccPIE}. Since this is difficult to achieve experimentally, some new developments in the probe position corrections have been made.    

The non-linear (NL) optimization approach was the first method that has been used to correct the probe positions \cite{nonlinear}. However, this approach can easily lead to local minima which can be far from the required global minimum since several parameters (update of the object, the probe function and the probe positions) are used in the NL optimization routine. Improvements have been made in the NL optimization approach by combining it with ePIE and difference map (DM) \cite{conjugate_PIE}. In this reference, the authors have used the ePIE and DM to update the object and the probe function, whereas the probe positions have been corrected using the NL optimization. One drawback of this method is that the probe positions can not be corrected to sub-pixel accuracy. Other methods based on the genetic algorithm and a drift-based model were also explored \cite{evolutionary,Drift}. In yet another study, the ``annealing approach" ``based on trial and error" was used, but at the cost of being computationally expensive \cite{pcPIE}. Finally, there is a successful method that uses the cross-correlation between two consecutive object estimates for each probe position \cite{ccPIE}. This approach has corrected the probe positions to sub-pixel accuracy using the additional sub-pixel registration method \cite{sub-pixel}.          

Here, we analyse and demonstrate an alternative algorithm to correct the probe positions with sub-pixel accuracy that is quite straightforward to implement \cite{SPIE}. This paper is organized as follows: In Section 2, we describe our method for the probe position correction. In Section 3, the robustness of the method will be verified by evaluating the simulation results. In Section 4, we show the experimental results. Finally, in Section 5, we present the conclusions.

\section{The algorithm}

In ptychography, the diffraction intensities $I^j(\mathbf{u})$ for different probe positions $j$ = 1,2,...,$J$ with respect to the object are recorded in the camera. Here, $J$ is the number of diffraction patterns. If the object and illumination probe functions are represented as $O(\mathbf{r})$ and $P(\mathbf{r})$, then 
\begin{equation}
I^j(\mathbf{u}) = |\mathscr{F}\{O(\mathbf{r})P(\mathbf{r}-\mathbf{R}^j)\}(\mathbf{u})|^2,
\end{equation}
where $\mathbf{R}^j = (X^j,Y^j)$ is the probe position vector, $\mathbf{\mathbf{r}}$ and $\mathbf{u}$ represent the coordinate vector in the real and reciprocal space respectively, and $\mathscr{F}$ denotes the Fourier transform. We combine the well-known phase reconstruction method ePIE with our position correction method. That means, for the $k^\text{th}$ iteration and the $j^\text{th}$ probe position, we update the object $O_k(\mathbf{r})$ to $O_{k+1}(\mathbf{r})$ and the probe function $P_k(\mathbf{r})$ to $P_{k+1}(\mathbf{r})$ using the ePIE after which the probe position $\mathbf{R}_k^j$ is updated using our probe position correction method. We describe the probe position correction method below. Note that in this probe position correction method, we use the previous estimates $O_k(\mathbf{r})$ and $P_k(\mathbf{r})$ instead of $O_{k+1}(\mathbf{r})$ and $P_{k+1}(\mathbf{r})$ as this saves one extra Fourier transform to perform. The reason will be clear soon.   

For the $k^{\text{th}}$ iteration, the diffracted far field for the probe position $\mathbf{R}_k^j$ can be written as 
\begin{equation}
\Psi_k^j(\mathbf{u}) = \mathscr{F}\{O_k(\mathbf{r})P_k(\mathbf{r}-\mathbf{R}_k^j)\},
\end{equation}
and the estimated intensity is 
\begin{equation}
I_{k}^j(\mathbf{u}) = |\Psi_k^j(\mathbf{u})|^2.
\end{equation}
For the object estimate $O_k(\mathbf{r})$ and probe estimate $P_k(\mathbf{r})$, the inaccuracy in the measurement intensity due to the error $(\Delta X_k^j, \Delta Y_k^j)$ in the probe position is given by: 
\begin{align}\label{dI}
\begin{split}
\Delta I_k^j &\approx \frac{\partial I_k^j}{\partial X_k^j} \Delta X_k^j + \frac{\partial I_k^j}{\partial Y_k^j} \Delta Y_k^j. \\
\end{split}
\end{align}
Here, $\frac{\partial I_k^j}{\partial X_k^j}$ and $\frac{\partial I_k^j}{\partial Y_k^j}$ are the derivatives of the estimated intensity with respect to the probe position along the $x$ and $y$ directions.
We solve Eq. \eqref{dI} for $\Delta X_k^j$ and $\Delta Y^j_k$ where $\Delta I_k^j$ is assigned to $I^j-I_{k}^j$. To calculate $\frac{\partial I^j_k}{\partial X^j_k}$ and $\frac{\partial I^j_k}{\partial Y^j_k}$, we have 
\begin{subequations} \label{dI_XY}
\begin{align}
 	\frac{\partial I^j_k}{\partial X^j_k} &= 2Re\bigg\{\frac{\partial \Psi^j_k}{\partial X^j_k}\Psi_k^{j*}\bigg\}, \\
 	\frac{\partial I^j_k}{\partial Y^j_k} &= 2Re\bigg\{\frac{\partial \Psi^j_k}{\partial Y^j_k}\Psi_k^{j*}\bigg\},
\end{align}
\end{subequations}
and 
\begin{subequations}\label{dPSI_XY1}
\begin{align}
\frac{\partial \Psi_k^j(\mathbf{u})}{\partial X_k^j} = \mathscr{F}\bigg\{O_k(\mathbf{r})\frac{\partial P_k(\mathbf{r}-\mathbf{R}_k^j)}{\partial X_k^j}\bigg\}(\mathbf{u}),
\\
\frac{\partial \Psi_k^j(\mathbf{u})}{\partial Y_k^j} = \mathscr{F}\bigg\{O_k(\mathbf{r})\frac{\partial P_k(\mathbf{r}-\mathbf{R}_k^j)}{\partial Y_k^j}\bigg\}(\mathbf{u}).
\end{align}
\end{subequations}
We approximate the right hand side of the Eq. \eqref{dPSI_XY1} as 
\begin{subequations}\label{dPsi_XY2}
 \begin{align}
 \frac{\partial \Psi^j_k}{\partial X^j_k} &= \frac{\Psi^j_k - \mathscr{F}\{O_{k}(\mathbf{r})P_{k}(\mathbf{r}-(\mathbf{R}^j_k+\mathbf{1}_x))\}}{|\mathbf{1}_x|}, 	\\
 \frac{\partial \Psi^j_k}{\partial Y^j_k} &= \frac{\Psi^j_k - \mathscr{F}\{O_{k}(\mathbf{r})P_{k}(\mathbf{r}-(\mathbf{R}^j_k+\mathbf{1}_y))\}}{|\mathbf{1}_y|},
 \end{align}
 where $\mathbf{1}_x$ and $\mathbf{1}_y$ are the vectors along the x and y directions and the magnitudes are the lengths of a pixel along the x and y directions, respectively.  
\end{subequations}
The following steps are performed to calculate the error and update the probe positions. 

\begin{enumerate} 
\item Calculate the difference $\Delta I_k^j$ between the measured intensity $I^j$ and the estimated intensity $I^j_{k}$ given by $\Delta I_k^j=I^j-I^j_{k}.$
\item Calculate $\frac{\partial \Psi^j_k}{\partial X^j_k}$ and $\frac{\partial \Psi^j_k}{\partial Y^j_k}$ using Eq. \eqref{dPsi_XY2}.
\item Calculate $\frac{\partial I^j_k}{\partial X^j_k}$ and $\frac{\partial I^j_k}{\partial Y^j_k}$ using Eqs. \eqref{dI_XY}. 

\end{enumerate}
Note that in Eq. \eqref{dI}, $\Delta I^j_k$, $\frac{\partial I^j_k}{\partial X^j_k}$, and $\frac{\partial I^j_k}{\partial Y^j_k}$ are vectors whose components correspond to the values at the pixels. Given an intensity measurement consisting of N pixels, we can thus rewrite Eq. \eqref{dI} as a matrix equation
\begin{equation}\label{Mat}
\begin{bmatrix}
\Delta I^j_k(1)
\\
\Delta I^j_k(2)
\\
\vdots
\\
\Delta I^j_k(N)
\end{bmatrix}
=
\begin{bmatrix}
\frac{\partial I^j_k}{\partial X^j_k} (1) & \frac{\partial I^j_k}{\partial Y^j_k}(1)
\\
\frac{\partial I^j_k}{\partial X^j_k} (2) & \frac{\partial I^j_k}{\partial Y^j_k}(2)
\\
\vdots
\\
\frac{\partial I^j_k}{\partial X^j_k} (N) & \frac{\partial I^j_k}{\partial Y^j_k}(N)
\end{bmatrix}
\begin{bmatrix}
\Delta X^j_k
\\
\Delta Y^j_k
\end{bmatrix}.
\end{equation}
From this equation, we want to find $(\Delta X^j_k, \Delta Y^j_k)$. Because there are more equations than variables, there may be no solution $(\Delta X^j_k, \Delta Y^j_k)$ to this equation. Therefore, we calculate the least-squares solution which is given by 
\begin{equation}
\begin{bmatrix}
\Delta X^j_k
\\
\Delta Y^j_k
\end{bmatrix}
=(A^{jT}_k A^j_k)^{-1} A^{jT}_k
\begin{bmatrix}
\Delta I^j_k(1)
\\
\Delta I^j_k(2)
\\
\vdots
\\
\Delta I^j_k(N)
\end{bmatrix},
\end{equation}
where
\begin{equation}
A^{j}_k = \begin{bmatrix}
\frac{\partial I^j_k}{\partial X^j_k} (1) & \frac{\partial I^j_k}{\partial Y^j_k}(1)
\\
\frac{\partial I^j_k}{\partial X^j_k} (2) & \frac{\partial I^j_k}{\partial Y^j_k}(2)
\\
\vdots
\\
\frac{\partial I^j_k}{\partial X^j_k} (N) & \frac{\partial I^j_k}{\partial Y^j_k}(N)
\end{bmatrix},
\end{equation}
and $A^{jT}_k$ is the transpose of $A^{j}_k$. Note that $A^{jT}_k A^j_k$ is a $2\times 2$ matrix, so the computation of its inversion is computationally inexpensive. Finally, the update equation for the probe position $(X^j_k,Y^j_k)$ is given by
\begin{align}
X^j_{k+1} &= X^j_{k}-\beta \Delta X^j_{k}, \\
Y^j_{k+1} &= Y^j_{k}-\beta \Delta Y^j_{k}.
\end{align}
Here, $\beta$ is a feedback parameter which defines the step size of the update in the probe positions. Choosing smaller $\beta$ in general leads to accurate correction but the computation time is larger. The value of $\beta$ can be chosen as 1, 0.5 or 0.1.

To compare our approach to the NL optimization method \cite{nonlinear} note that the cost function used in the NL optimization integrates over all pixels, whereas our optimization approach considers the change for each pixel. In other words, the cost function of the NL optimization can have the same value for different configuration of pixel values whereas our optimization will not. On comparing the computational time of our approach with the cross-correlation (CC) method \cite{ccPIE}, we have found that each iteration of our method is less computationally expensive than CC method. Here, in the probe position correction part, we are using two Fourier transforms whereas the CC method uses three Fourier transforms. Additionally, the CC method requires an optimization in each iteration to find the cross-correlation peak. Furthermore, in the section 3.4, we have carried out an actual comparison between CC method and proposed method.         
 
\section{Simulations}
\subsection{Simulations on the general performance of the algorithm}
We denote the wavelength of the light by $\lambda$, the detector pixel size by $\Delta x_d=\Delta y_d$, the distance between the object and the detector by $z$. The detector has $N$ pixels and the number of pixels in the x and y directions are the same, i.e. both are $\sqrt{N}$. If the detector pixel size is $\Delta x_d$, the pixel size in the Fourier space is $\Delta f = \frac{k\Delta x_d}{z}$. The pixel size in the real space (i.e. the object space) $\Delta x_o$ can be found by using the relation $\Delta x_o\Delta f = \frac{2\pi}{\sqrt{N}}$. Thus, $\Delta x_o = \frac{\lambda z}{\sqrt{N} \Delta x_d}$.  

We used a circular illumination probe which was formed by propagating an uniformly illuminated circular pinhole function with a diameter of $89.6\Delta x_o$. We have used a gray probe which means that a pixel can have a non-integer value between $0$ and $1$; consequently, the diameter of the probe is represented with sub-pixel accuracy. The propagation distance $z'$ is chosen such that $\lambda z' = 5\times 10^{-4}\text{mm}^2 $. The size of the scanned object is $224\Delta x_o \times 224\Delta x_o$. The probe positions were formed using a grid of $8\times 8$. The grid interval was $19.2\Delta x_o$, and the overlap between the adjacent probes was $73\%$. Random offsets with a maximum value of $10\Delta x_o$, were added to each probe position in both x and y directions. These generated probe positions were used to form the far-field intensity patterns. The feedback parameter $\beta$ was chosen to be 0.5.    

`Cameraman' was used as the test object with amplitude which varies between $[0,1]$. `Lenna' was used as the phase of the test object with values from $[-0.7\pi,0.7\pi]$. In our simulation, the probe position update starts at $15^{\text{th}}$ iterations and probe function update starts at $45^{\text{th}}$ iterations. The simulation ran for 300 iterations. Figs. 1(a-d) show the object and probe functions which were used to generate the simulated diffraction patterns. Figs. 1(e-h) show the reconstruction of the object and probe functions using the ePIE when the error in the probe positions was present. Figs. 1(i-l) show the reconstruction of the object and probe when our approach to correct the probe positions together with the ePIE was used. Note that the contours of the object amplitude are visible in the reference phase (Fig. 1(c)) and reconstructed phase (Fig. 1(k)). These are due to the presence of zero amplitude in the object.   

In Fig.~\ref{scatter}, we map the updates of the probe positions as they converge from the initial guessed positions to the actual probe positions. The green dots represent the actual probe positions which were used to generate the intensity patterns in the far-field, the red dots are the initial guesses for the probe positions, and the blue dots indicate the trajectory of the convergence. Note that almost all initially guessed positions converge to the actual positions. In Fig. \ref{diff_er}, the plot for diffraction error versus iteration is shown. The error metric for each iteration is defined as
\begin{equation}
\mathcal{E}^j = \sum_\mathbf{u} \left\{|\Psi_k^j(\mathbf{u})| - \sqrt{I^j(\mathbf{u})}\right\}^2
\label{error}
\end{equation}
Fig. \ref{diff_er} shows the mean of $\mathcal{E}^j$ over all the probe positions for each iteration.   

\begin{figure}[H]
        \centering

		\begin{subfigure}[]{0.3\textwidth}
            \centering
            \textbf{Object Amplitude}
            \includegraphics[width=\textwidth]{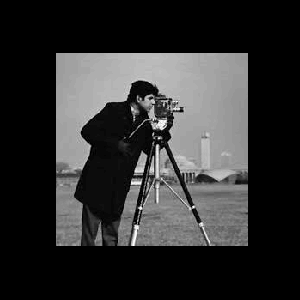}
            \caption{}%
        \end{subfigure}  
        \begin{subfigure}[]{0.15\textwidth}
            \centering
            \includegraphics[width=\textwidth]{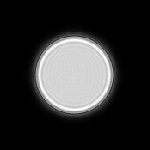}
            \caption{}%
        \end{subfigure}  
        \begin{subfigure}[]{0.3\textwidth}
            \centering
            \textbf{Object Phase}
            \includegraphics[width=\textwidth]{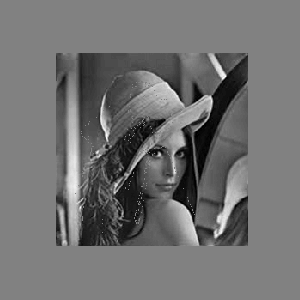}
            \caption{}%
        \end{subfigure}
        \begin{subfigure}[]{0.15\textwidth}
            \centering
            \includegraphics[width=\textwidth]{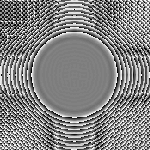}
            \caption{}%
        \end{subfigure}
        
		\begin{subfigure}[H]{0.3\textwidth}  
            \centering 
            \includegraphics[width=\textwidth]{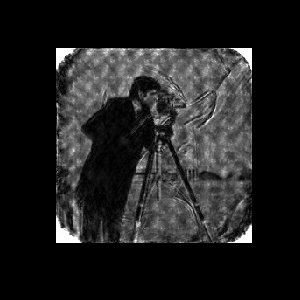}
            \caption{}%
        \end{subfigure}
        \begin{subfigure}[]{0.15\textwidth}
            \centering
            \includegraphics[width=\textwidth]{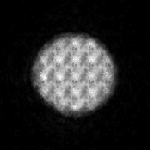}
            \caption{}%
        \end{subfigure}
        \begin{subfigure}[H]{0.3\textwidth}  
            \centering 
            \includegraphics[width=\textwidth]{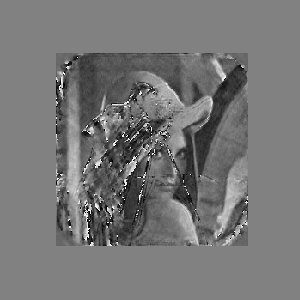}
            \caption{}%
        \end{subfigure}
        \begin{subfigure}[]{0.15\textwidth}
            \centering
            \includegraphics[width=\textwidth]{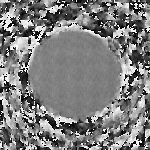}
            \caption{}%
        \end{subfigure}        
        
        \begin{subfigure}[]{0.3\textwidth}
            \centering
            \includegraphics[width=\textwidth]{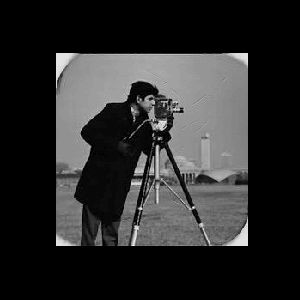}
            \caption{}%
        \end{subfigure}  
        \begin{subfigure}[]{0.15\textwidth}
            \centering
            \includegraphics[width=\textwidth]{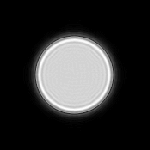}
            \caption{}%
        \end{subfigure}  
        \begin{subfigure}[]{0.3\textwidth}
            \centering
            \includegraphics[width=\textwidth]{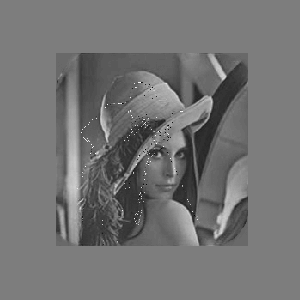}
            \caption{}%
        \end{subfigure}
        \begin{subfigure}[]{0.15\textwidth}
            \centering
            \includegraphics[width=\textwidth]{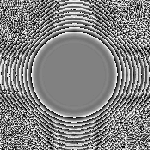}
            \caption{}%
        \end{subfigure}

        \caption{Reconstruction of the object and probe after 300 iterations. (a)-(d) are the test object and probe functions (Figs. (b) and (d)), which were used to generate the diffraction intensity patterns. The object amplitude varies between [0,1], and the object phase ranges from [-0.7$\pi$,0.7$\pi$]. (e)-(h) are the reconstructed images using the ePIE. (i)-(l) are the reconstructed images using the ePIE together with the probe positions correction. For these simulations, $8\times8$ probe positions with an overlap of 73\% have been used.}
\end{figure}

\begin{figure}[H]

	\centering
	\includegraphics[width=0.6\textwidth]{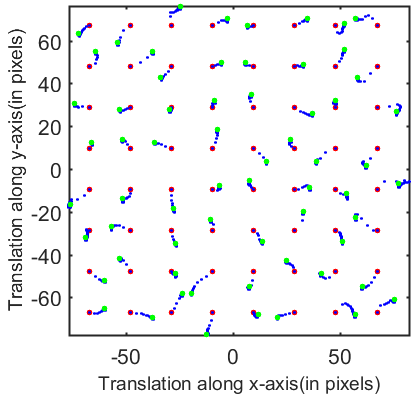}
	\caption{Probe position correction map. Here, the maximum random added offset to each probe position is $10\Delta x_o$. The red dots are the initial guess for the probe positions, the green dots are the actual probe positions, and the blue dots show how the estimated probe positions change over iterations.}
		\label{scatter}
\end{figure}

\begin{figure}[H]
\centering 
\includegraphics[width=0.75\textwidth]{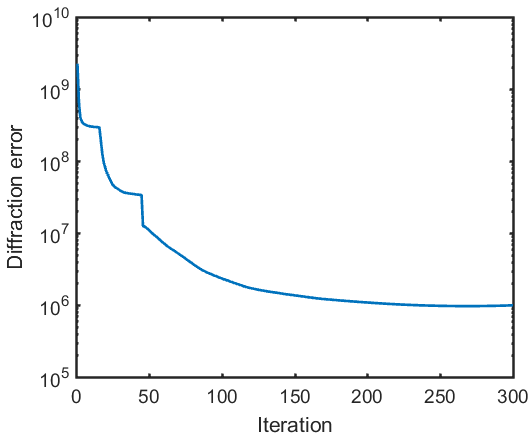}
\caption{Plot for diffraction error versus iteration. The diffraction errors for each probe position is calculated using Eq. \eqref{error}. The plot is mean of diffraction error for all the probe positions versus iteration.}
\label{diff_er}
\end{figure}

\subsection{Simulations in the presence of noise}
\begin{figure}[H]
        \centering

		\begin{subfigure}[]{\textwidth}
            \centering
            \includegraphics[width=0.75\textwidth]{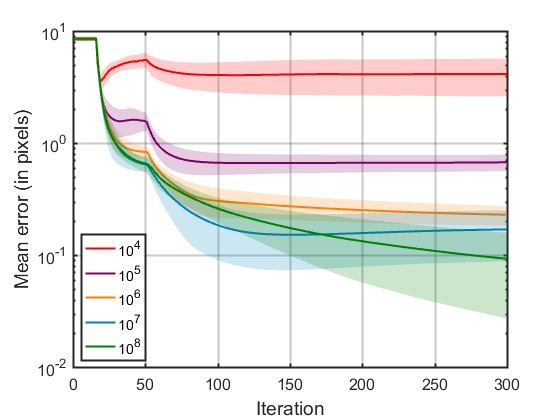}
            \caption{}%
        \end{subfigure}  
       \caption{Mean error in the probe position in the presence of Poisson noise. The solid lines represent the mean value, whereas the patches show the standard deviation. For each noise level (varying from $10^8$ to $10^4$ photons per diffraction pattern), ten simulations were made where different random initial offsets are used to generate the actual probe positions.}
       \label{noise}
\end{figure}

The algorithm was also tested in the presence of Poisson noise. We performed simulations with different numbers of photons per diffraction pattern. The simulation for each noise level was run for ten times with different random initial offsets taken from [-10,10] pixels along the x and y directions. The other parameters of the object and probe functions were the same as in section 3.1. In fig. \ref{noise}, the solid lines represent the mean value whereas the patches show the standard deviation. If $\sigma_j$ represents a standard deviation over all j, then the mean error is calculated as following: 
\begin{equation}
E_k = \sqrt{\big\{ \sigma_j(X^j-X^j_k)\big\}^2+\big\{\sigma_j(Y^j-Y^j_k)\big\}^2}.
\end{equation}

Even with approximately $10^5$ photons per diffraction pattern, the mean error in the retrieved probe positions was less than one pixel, and the mean error for the case of $10^8$ photons was as low as $10^{-2}$ pixel.    

\subsection{Effect of overlap and initial position error}
\begin{figure}[H]
        \centering
		\begin{subfigure}[]{.48\textwidth}
            \centering
            \includegraphics[width=1\textwidth]{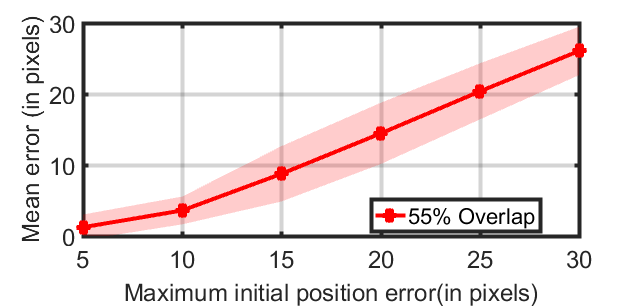}
            \caption{}%
        \end{subfigure} 
        \hfill 
        \begin{subfigure}[]{.48\textwidth}
            \centering
            \includegraphics[width=1\textwidth]{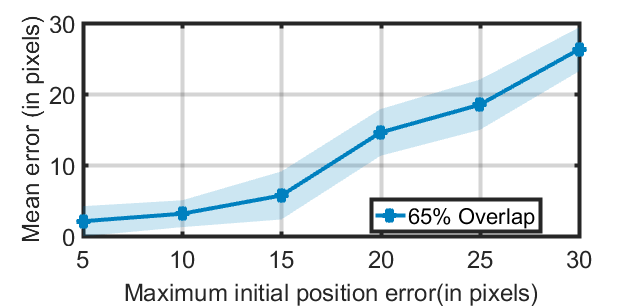}
            \caption{}%
        \end{subfigure} 
        \hfill
        \begin{subfigure}[]{.48\textwidth}
            \centering
            \includegraphics[width=1\textwidth]{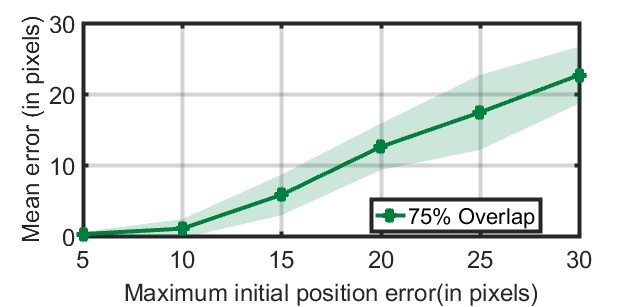}
            \caption{}%
        \end{subfigure}  
        \hfill
        \begin{subfigure}[]{.48\textwidth}
            \centering
            \includegraphics[width=1\textwidth]{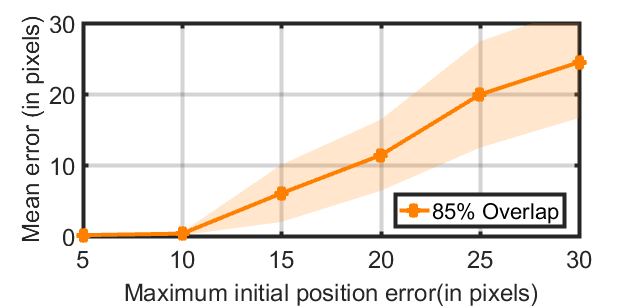}
            \caption{}%
        \end{subfigure}   
       \caption{Results for varying overlaps and initial position errors versus mean error in the probe position. For each overlap and introduced maximum initial position error, ten simulations were performed with different random initial offsets. Here, the solid lines represent the mean value of the final position error of the ten simulations, whereas the patches show the standard deviation.}
       \label{overlap}
\end{figure}

In Fig.\ref{overlap}, the performance of the algorithm for different overlaps and different introduced initial position errors have been shown. For each overlap and maximum introduced initial position error, ten simulations were performed with different random initial offsets. The solid lines represent the mean value whereas the patches show the standard deviation. $6 \times 6$ probe positions have been used for the results shown here. Since the final error is gradually increasing as the initial introduced maximum error is increasing, it is difficult to comment on  the maximum initial error this method can correct for this case. There is also no specific point where a sharp increase in the error can be seen. The similar behaviour is also observed in the Ref. \cite{conjugate_PIE}. From Fig. \ref{overlap}, the correction of the probe positions is not strongly dependent on overlap. However, $75\%$ overlap can be considered as optimum overlap for this case.

\subsection{Comparison with cross-correlation method}
\begin{figure}[H]
\centering
	\begin{subfigure}{0.48\textwidth}
	\centering
	\includegraphics[width=\textwidth]{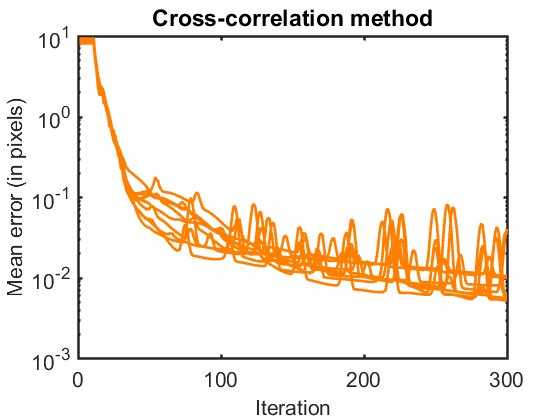}
	\caption{}
	\end{subfigure}
		\begin{subfigure}{0.48\textwidth}
	\centering
	\includegraphics[width=\textwidth]{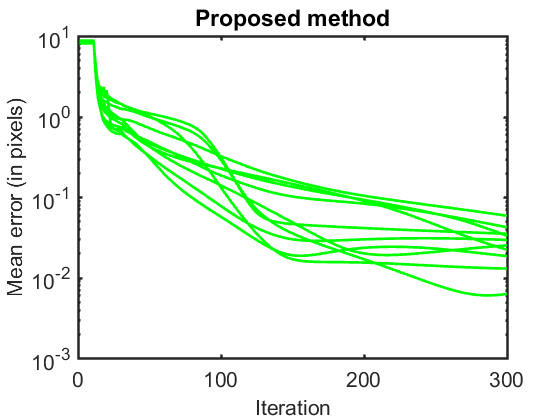}
	\caption{}
	\end{subfigure}
	\caption{Comparison between the cross-correlation method and proposed method. For each method, ten simulations were performed where the random initial offsets were used to generate the actual probe positions. (a) Mean error in the probe position versus iteration for the cross-correlation method. (b) Mean error in the probe position versus iteration for the proposed method.}
	\label{comparison}
\end{figure}

It was previously noted that an iteration of the proposed method is less computationally expensive than an iteration of the cross-correlation (CC) method. However, this does not necessarily mean that the proposed method is less computationally expensive on the overall, because in principle it could require more iterations to obtain the same reconstruction error. Therefore, we compare the two methods to see if indeed the proposed method is less computationally expensive than the CC method. Here, we have performed ten simulations for each method where the parameters are same as given in the simulation section and the random initial offsets for probe positions were taken from $[-10,10]$ pixels. In Fig. \ref{comparison}(a), we have encountered the small bumps in the simulation which are due to the implementation of automatic adapting feedback parameter as explained in the Ref. \cite{ccPIE}. Here, threshold parameters for the automatic feedback parameters are $-0.2$ and $0.45$. In Fig. \ref{comparison}(b), simulations for the proposed method is shown where the feedback parameter is $1$. The final mean error of these ten simulations after $300$ iterations for the CC method and the proposed method are $0.013$ and $0.023$ pixels respectively. From here, we draw the conclusion that even though, initially, the CC method converges faster than proposed method, both methods achieve comparable accuracy once they converge.

\section{Experiment}
To demonstrate the algorithm to correct the probe positions, we built an experimental setup as shown in Fig. 7. The beam of a HeNe laser (633 nm wavelength) is first expanded and collimated by two lenses and used to illuminate the phase-only Spatial light modulator (Holoeye Pluto, 1920 $\times$ 1080 pixels, 8.0 $\mu$m pixel pitch). In the SLM, a phase pattern is created (Lenna, with the phase varying from 0 to 1.8 $\pi$) together with an added phase ramp to shift the non-modulated from the modulated signals at the diffraction plane [17]. This created phase pattern on the SLM is the assigned image that we want to reconstruct. The illumination probe was created by adding a rapidly phase changing pattern (Fig. \ref{ex_recons_2} c) with SLM on top of the object. The far-field was obtained by using a 15 cm focal length lens placed at 15 cm from the SLM. The CCD camera (8 bit, pixel size 4.65 $\mu\text{m}$) was placed at the back focal plane of the lens to collect the ptychographic data set. On the SLM, the object had 800 $\times$ 800 pixels, and it was illuminated by a circle of radius of 250 pixels. The object was shifted to 7 $\times$7 positions with an interval of 50 pixels which is equivalent to shift the probe in reverse order. Due to magnification, the added random offset in simulation of $5\Delta x_o$ is equivalent to 20.83 pixels on SLM.

In Figs. \ref{ex_recons_2}-\ref{ex_recons_10}, a comparison has been shown between the ePIE and ePIE with our probe position correction method for different added random offsets to the probe positions. Maximum random offsets added to the probe positions in the simulation were $2\Delta x_o$ and $10\Delta x_o$ respectively. In Fig. \ref{ex_scatter}, the scatter plot is shown when the maximum initial random offset was $2\Delta x_o$. Fig. \ref{ex_scatter}(a) shows the actual probe positions (green) and initial guessed probe positions (red) whereas Fig. \ref{ex_scatter}(b) shows the actual probe positions (green) and the final reconstructed probe positions (red). Note that the final reconstructed probe positions shown here are translated by a constant. In Fig. \ref{ex_EOD}(a), the plot for diffraction error versus introduced maximum initial position error is shown. The diffraction error is calculated using Eq. \eqref{error}.  To show what it means to have a diffraction error of $10^6$, in Fig. \ref{ex_EOD}(b), the estimated amplitude, measured amplitude, and its difference is shown for the probe position (1,1).

\begin{figure}[H]
	\centering
	\includegraphics[width=0.7\textwidth]{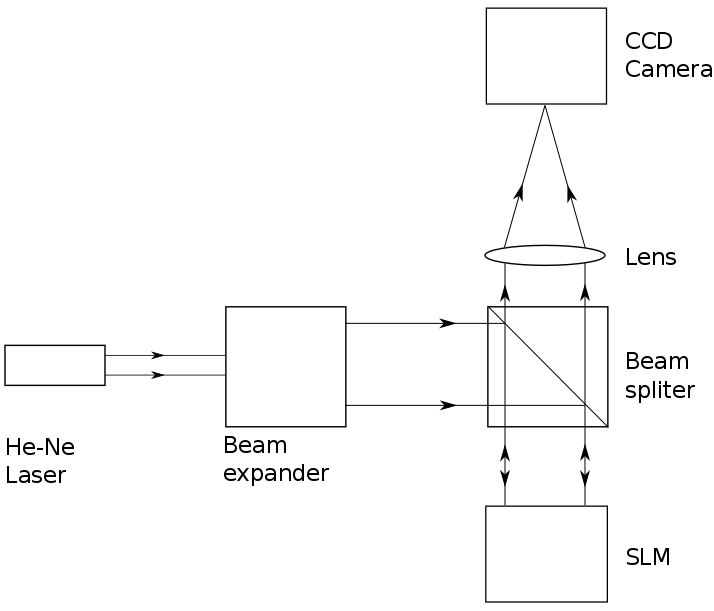}
	\caption{Experimental set-up. The set-up consists of He-Ne Laser (633 nm wavelength), beam expander, beam spliter, lens (f = 15cm), SLM (Holoeye Pluto, 1920$\times$1080 pixels, 8.0 $\mu$m pixel pitch), and CCD Camera (8 bit, pixel size 4.65 $\mu$m$^2$).}
\label{set_up}
\end{figure}
\begin{figure}[H]
        \centering
		\begin{subfigure}[t]{0.21\textwidth}
            \centering
            \includegraphics[width=\textwidth]{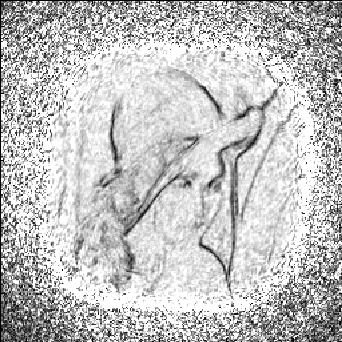}
            \subcaption{ePIE, amplitude}
        \end{subfigure} 
        \hfill
     	\begin{subfigure}[t]{0.21\textwidth}
        \centering       
            \includegraphics[width=\textwidth]{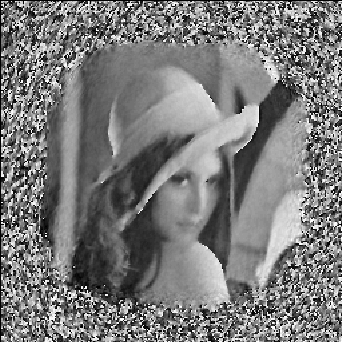}
 		\subcaption{Phase}%
        \end{subfigure}
        \hfill
        \begin{subfigure}[t]{0.21\textwidth}
            \centering
            \includegraphics[width=\textwidth]{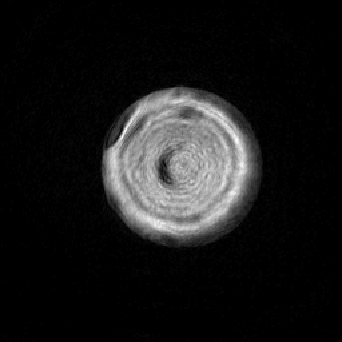}            
            \caption{Amplitude}%
        \end{subfigure}
        \hfill
        \begin{subfigure}[t]{0.21\textwidth}
            \centering
            \includegraphics[width=\textwidth]{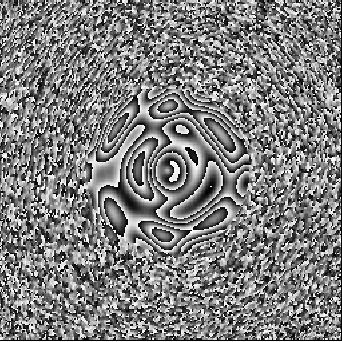}
            \caption{Phase}%
        \end{subfigure}
		\hfill
		\begin{subfigure}[t]{0.21\textwidth}
            \centering
            \includegraphics[width=\textwidth]{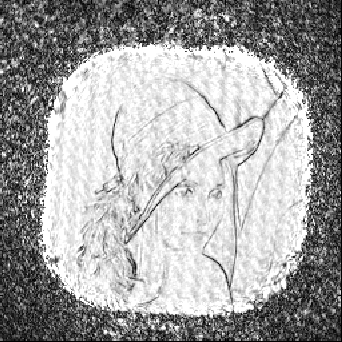}
            \caption{ePIE with positions correction, amplitude}
        \end{subfigure} 
        \hfill
     	\begin{subfigure}[t]{0.21\textwidth}
            \centering
            \includegraphics[width=\textwidth]{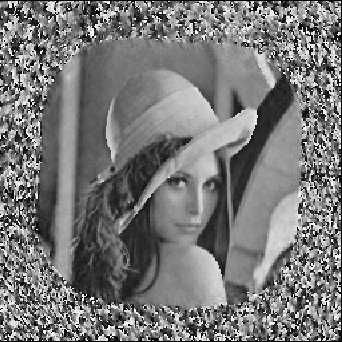}
            \caption{Phase}%
        \end{subfigure}
        \hfill
        \begin{subfigure}[t]{0.21\textwidth}
            \centering
            \includegraphics[width=\textwidth]{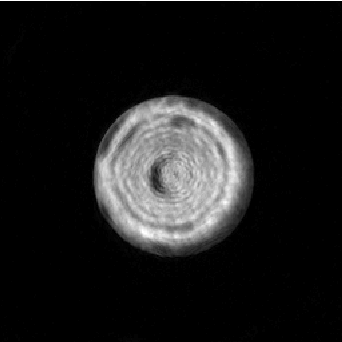}
            \caption{Amplitude}%
        \end{subfigure}
        \hfill
        \begin{subfigure}[t]{0.21\textwidth}
            \centering
            \includegraphics[width=\textwidth]{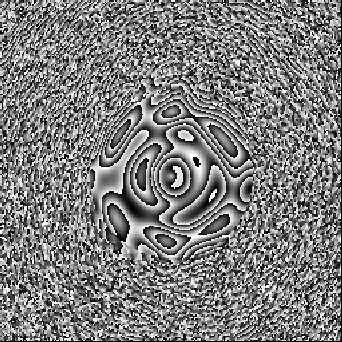}
            \caption{Phase}%
        \end{subfigure}
\caption{Maximum random added offset to each probe position = $2\Delta x_o$. We have truncated the color scale for the object amplitude to make certain features more visible.}
\label{ex_recons_2}
\end{figure}      

\begin{figure}[H]
		\begin{subfigure}[t]{0.21\textwidth}
            \centering
            \includegraphics[width=\textwidth]{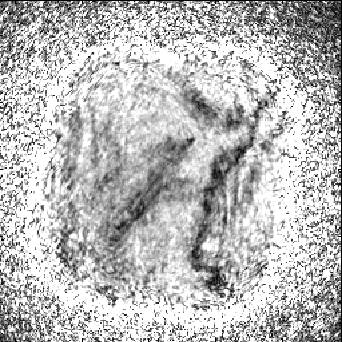}
            \subcaption{ePIE, amplitude}
        \end{subfigure} 
        \hfill
     	\begin{subfigure}[t]{0.21\textwidth}
            \centering       
            \includegraphics[width=\textwidth]{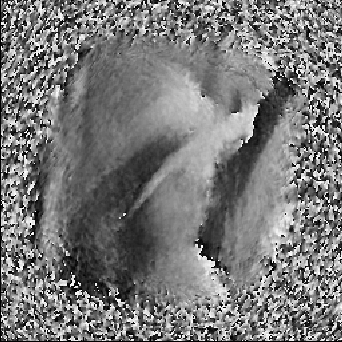}
        \subcaption{Phase}%
        \end{subfigure}
        \hfill
        \begin{subfigure}[t]{0.21\textwidth}
            \centering
            \includegraphics[width=\textwidth]{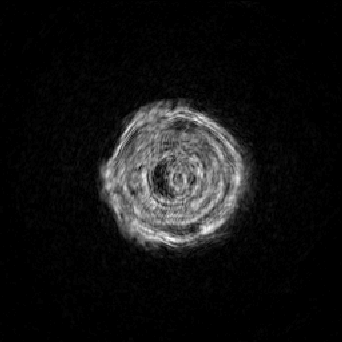}            
            \caption{Amplitude}%
        \end{subfigure}
        \hfill
        \begin{subfigure}[t]{0.21\textwidth}
            \centering
            \includegraphics[width=\textwidth]{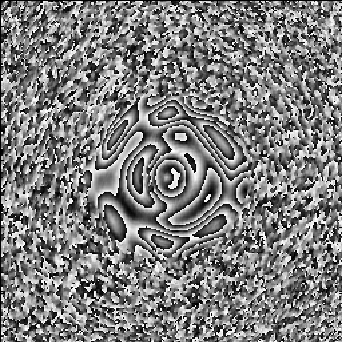}
            \caption{Phase}%
        \end{subfigure}
        \hfill
		\begin{subfigure}[t]{0.21\textwidth}
            \centering
            \includegraphics[width=\textwidth]{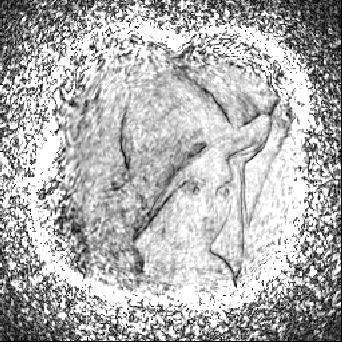}
            \caption{ePIE with position correction, amplitude}
        \end{subfigure} 
        \hfill
     	\begin{subfigure}[t]{0.21\textwidth}
            \centering
            \includegraphics[width=\textwidth]{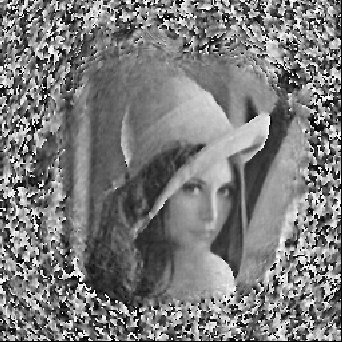}
            \caption{Phase}%
        \end{subfigure}
        \hfill
        \begin{subfigure}[t]{0.21\textwidth}
            \centering
            \includegraphics[width=\textwidth]{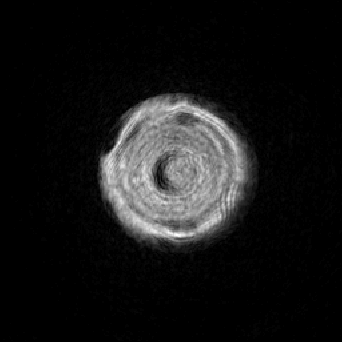}
            \caption{Amplitude}%
        \end{subfigure}
        \hfill
        \begin{subfigure}[t]{0.21\textwidth}
            \centering
            \includegraphics[width=\textwidth]{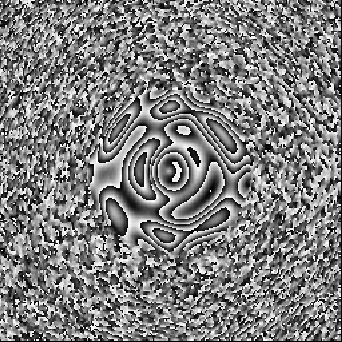}						
            \caption{Phase}%
        \end{subfigure} 
\caption{Maximum random added offset to each probe position = $10\Delta x_o$. We have truncated the color scale for the object amplitude to make certain features more visible.}
\label{ex_recons_10}
\end{figure}

\begin{figure}[H]
\centering
\includegraphics[width=\textwidth]{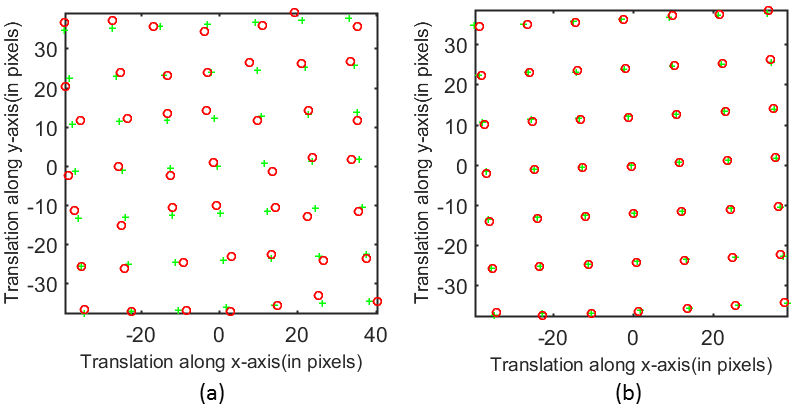}
\caption{Scatter plot of the probe positions are shown for the case when the maximum initial random offset added to each probe position was $2\Delta x_o$. The green marks are the actual probe positions. (a) The red circles represent the initial guessed probe positions. (b) The red circles represent the final reconstructed probe positions.}
\label{ex_scatter}
\end{figure}

\begin{figure}[H]
	\begin{subfigure}{\textwidth}
		\centering
		\includegraphics[width=0.7\textwidth]{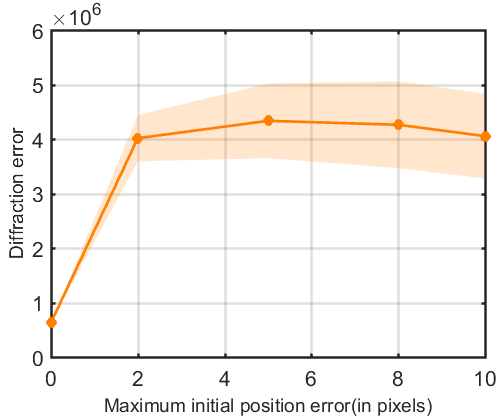}
		\caption{}
	\end{subfigure}
	\begin{subfigure}{\textwidth}
		\includegraphics[width=\textwidth]{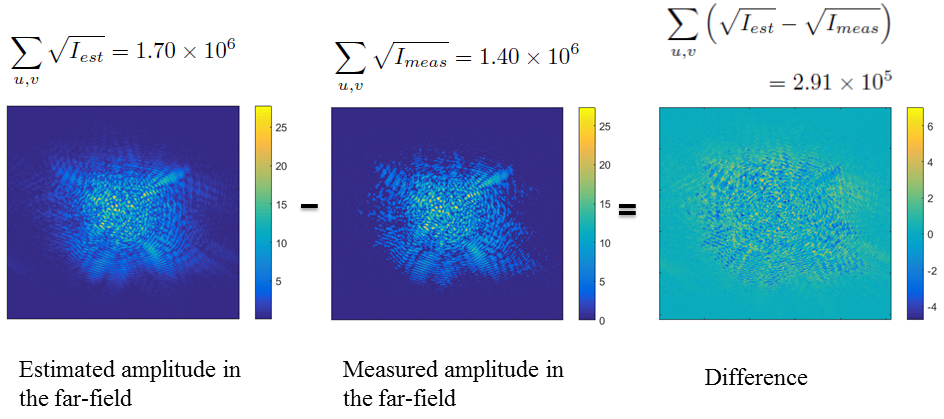}
		\caption{}
	\end{subfigure}
\caption{Diffraction error in the far-field. (a) Here, for each maximum initial position error, ten simulations were performed with different random initial offsets. The solid line represents the mean value of the final diffraction error of the ten simulations, whereas the patch shows the standard deviation. The diffraction error is calculated using Eq. \eqref{error}.(b) To show what it means to have diffraction error of $10^6$, estimated, measured amplitude in the far-field and its difference are shown here for the probe position $(1,1)$ and when there is no error in the probe position.}
\label{ex_EOD}
\end{figure}

\section{Conclusion}

We have tested a novel technique to correct the lateral probe positions in ptychography. It is a straightforward extension to the ePIE, and with simulations, we showed that it can correct the probe positions to sub-pixel accuracy even in the presence of noise. Each iteration of this method is less computationally inexpensive than the cross-correlation method and it achieves comparable accuracy once it is converged. Visible light experimental data was used to analyse this technique. Experimental results show significant improvements in the reconstruction. We anticipate that these results can be employed in realizing the full potential of ptychographic coherent diffractive imaging for high-resolution imaging.  

\section*{Acknowledgement}
The research leading to these results has received funding from the people programme (Marie Curie Actions) of the European Union's Seventh Framework Programme (FP72007-2013) under REA Grant Agreement no. PITN-GA-2013-608082.

\end{document}